\documentclass[5p,number,times,twocolumn]{elsarticle}
\usepackage[utf8]{inputenc}
\usepackage{float}
\usepackage{graphicx}
\usepackage{amsmath,amsthm,amssymb}
\usepackage{verbatim}
\usepackage{url}
\usepackage{subcaption}
\usepackage[separate-uncertainty=true]{siunitx}
\usepackage{physics}
\usepackage[colorlinks]{hyperref}
\usepackage[capitalize]{cleveref}
\usepackage{multirow}
\usepackage{orcidlink}
\usepackage{adjustbox}
\usepackage{cmap}
\biboptions{sort&compress}
\graphicspath{{figures/}}
\DeclareSIUnit\barn{b}

\journal{Physics Letters B}
\begin{document}

\title{Measurement of \texorpdfstring{$J/\psi$}{J/ψ} and \texorpdfstring{$\psi\left(2S\right)$}{ψ(2S)} production
	in \texorpdfstring{$p+p$}{p+p} and \texorpdfstring{$p+d$}{p+d} interactions at \texorpdfstring{\SI{120}{\GeV}}{120~GeV}}
\affiliation[UIUC]{
	organization={Department of Physics, University of Illinois at Urbana-Champaign},
	city={Urbana},
	state={Illinois 61801},
	country={USA}}
\affiliation[Tokyo]{
	organization={Department of Physics, Tokyo Institute of Technology},
	city={Meguro-ku},
	state={Tokyo 152-8550},
	country={Japan}}
\affiliation[AS]{
	organization={Institute of Physics, Academia Sinica},
	city={Taipei 11529},
	country={Taiwan}}
\affiliation[LANL]{
	organization={Physics Division, Los Alamos National Laboratory},
	city={Los Alamos},
	state={New Mexico 87545},
	country={USA}}
\affiliation[RIKEN]{
	organization={RIKEN Nishina Center for Accelerator-Based Science},
	city={Wako},
	state={Saitama 351-0198},
	country={Japan}}
\affiliation[UVa]{
	organization={University of Virginia},
	city={Charlottesville},
	state={Virginia 22904},
	country={USA}}
\affiliation[NMSU]{
	organization={Department of Physics, New Mexico State University},
	city={Las Cruces},
	state={New Mexico 88003},
	country={USA}}
\affiliation[ANL]{
	organization={Physics Division, Argonne National Laboratory},
	city={Lemont},
	state={Illinois 60439},
	country={USA}}
\affiliation[Michigan]{
	organization={Randall Laboratory of Physics, University of Michigan},
	city={Ann Arbor},
	state={Michigan 48109},
	country={USA}}
\affiliation[ACU]{
	organization={Department of Engineering and Physics, Abilene Christian University},
	city={Abilene},
	state={Texas 79699},
	country={USA}}
\affiliation[FNAL]{
	organization={Fermi National Accelerator Laboratory},
	city={Batavia},
	state={Illinois 60510},
	country={USA}}
\affiliation[msstate]{
	organization={Department of Physics and Astronomy, Mississippi State University},
	city={Mississippi State},
	state={Mississippi 39762},
	country={USA} }
\affiliation[Rutgers]{
	organization={Department of Physics and Astronomy, Rutgers, The State University of New Jersey},
	city={Piscataway},
	state={New Jersey 08854},
	country={USA}}
\affiliation[Normal]{
	organization={Department of Physics, National Kaohsiung Normal University},
	city={Kaohsiung City 80201},
	country={Taiwan}}
\affiliation[Colorado]{
	organization={Department of Physics, University of Colorado},
	city={Boulder},
	state={Colorado 80309},
	country={USA}}
\affiliation[Yamagata]{
	organization={Department of Physics, Yamagata University},
	city={Yamagata City},
	state={Yamagata 990-8560},
	country={Japan}}
\affiliation[Maryland]{
	organization={Department of Physics, University of Maryland},
	city={College Park},
	state={Maryland 20742},
	country={USA}}
\affiliation[KEK]{
	organization={Institute of Particle and Nuclear Studies, KEK, High Energy Accelerator Research Organization},
	city={Tsukuba},
	state={Ibaraki 305-0801},
	country={Japan}}

\author[UIUC]{C.~H.~Leung\,\orcidlink{0000-0001-7907-3728}\fnref{fn_jlab}}
\author[Tokyo,AS,LANL]{K.~Nagai\,\orcidlink{0000-0002-5336-8306}\fnref{fn_duke}}
\author[Tokyo,RIKEN,UVa]{K.~Nakano\,\orcidlink{0000-0002-8925-2233}}
\author[NMSU]{D.~Nawarathne\,\orcidlink{0000-0001-5395-1190}}
\author[UIUC]{J.~Dove}
\author[UIUC,ANL]{S.~Prasad\,\orcidlink{0000-0003-3404-0062}}
\author[Michigan]{N.~Wuerfel\,\orcidlink{0000-0001-9872-5330}}

\author[Michigan,LANL]{C.~A.~Aidala\,\orcidlink{0000-0001-9540-4988}}
\author[ANL]{J.~Arrington\,\orcidlink{0000-0002-0702-1328}\fnref{fn_lbnl}}
\author[Michigan]{C.~Ayuso}
\author[ACU]{C.~L.~Barker}
\author[FNAL]{C.~N.~Brown}
\author[AS]{W.~C.~Chang\,\orcidlink{0000-0002-1695-7830}}
\author[UIUC,AS,Michigan]{A.~Chen}
\author[FNAL]{D.~C.~Christian\,\orcidlink{0000-0003-1275-6510}}
\author[UIUC]{B.~P.~Dannowitz}
\author[ACU]{M.~Daugherity\,\orcidlink{0000-0002-1781-9077}}
\author[msstate,Rutgers]{L.~El~Fassi\,\orcidlink{0000-0003-3647-3136}}
\author[ANL]{D.~F.~Geesaman}
\author[Rutgers]{R.~Gilman\,\orcidlink{0000-0002-7106-2845}}
\author[RIKEN]{Y.~Goto\,\orcidlink{0000-0002-2973-7458}}
\author[Normal]{R.~Guo}
\author[ACU]{T.~J.~Hague\,\orcidlink{0000-0003-1288-4045}\fnref{fn_lbnl}}
\author[ANL]{R.~J.~Holt\fnref{fn_caltech}\,\orcidlink{0000-0001-9225-9914}}
\author[NMSU]{M.~F.~Hossain\,\orcidlink{0000-0002-6467-1394 }}
\author[ACU]{D.~Isenhower\,\orcidlink{0000-0002-8237-5636}}
\author[Colorado]{E.~Kinney\,\orcidlink{0000-0002-4176-5283}}
\author[LANL]{A.~Klein}
\author[LANL]{D.~W.~Kleinjan\,\orcidlink{0000-0002-2737-0859}}
\author[Yamagata]{Y.~Kudo}
\author[Colorado,AS]{P.-J.~Lin\fnref{fn_ncu}}
\author[LANL]{K.~Liu\,\orcidlink{0000-0002-6676-8165}}
\author[LANL]{M.~X.~Liu\,\orcidlink{0000-0002-5992-1221}}
\author[Michigan]{W.~Lorenzon\,\orcidlink{0000-0003-0657-8463}}
\author[UIUC]{R.~E.~McClellan\fnref{fn_pensacola}}
\author[LANL]{P.~L.~McGaughey}
\author[ANL]{M.~M.~Medeiros}
\author[Yamagata]{Y.~Miyachi\,\orcidlink{0000-0002-8502-3177}}
\author[Tokyo]{S.~Miyasaka\,\orcidlink{0009-0004-1293-5679}}
\author[Michigan]{D.~H.~Morton\,\orcidlink{0000-0003-3813-1375}}
\author[Maryland]{K.~Nakahara\fnref{fn_slac}}
\author[Yamagata]{S.~Nara}
\author[NMSU]{S.~F.~Pate\,\orcidlink{0000-0001-8577-3405}}
\author[UIUC]{J.~C.~Peng\,\orcidlink{0000-0003-4198-9030}}
\author[NMSU]{A.~Pun}
\author[Michigan,FNAL]{B.~J.~Ramson\,\orcidlink{0000-0002-0925-3405}}
\author[ANL]{P.~E.~Reimer\,\orcidlink{0000-0002-0301-2176}}
\author[Michigan,ANL]{J.~G.~Rubin\,\orcidlink{0000-0002-9408-297X}}
\author[Tokyo]{F.~Sanftl}
\author[KEK]{S.~Sawada\,\orcidlink{0000-0002-7122-1690}}
\author[Michigan]{T.~Sawada\,\orcidlink{0000-0001-5726-7150}\fnref{fn_kagra}}
\author[Michigan,ANL]{M.~B.~C.~Scott\,\orcidlink{0000-0003-1105-1033}\fnref{fn_gwu}}
\author[Tokyo,RIKEN]{T.-A.~Shibata\,\orcidlink{0009-0005-5498-4804}\fnref{fn_nihon}}
\author[Rutgers]{A.~S.~Tadepalli\fnref{fn_jlab}}
\author[UIUC]{M.~Teo}
\author[ACU]{R.~S.~Towell\,\orcidlink{0000-0003-3640-7008}}
\author[LANL]{S.~Uemura\,\orcidlink{0000-0003-3458-4625}\fnref{fn_fnal}}
\author[AS,Normal]{S.~G.~Wang\,\orcidlink{0000-0001-8474-9817}\fnref{fn_aps}}
\author[LANL]{A.~B.~Wickes}
\author[FNAL]{J.~Wu}
\author[ANL]{Z.~H.~Ye\,\orcidlink{0000-0002-1873-2344}\fnref{fn_tsinghua}}

\fntext[fn_jlab]{Present address: Thomas Jefferson National Accelerator Facility, Newport News, Virginia 23606, USA.}
\fntext[fn_duke]{Present address: Duke University, Durham, North Carolina 27710, USA.}
\fntext[fn_lbnl]{Present address: Lawrence Berkeley National Laboratory, Berkeley, California, 94720 USA.}
\fntext[fn_caltech]{Present address: Kellogg Radiation Laboratory, California Institute of Technology, Pasadena, California 91125, USA.}
\fntext[fn_ncu]{Present address: Center for High Energy and High Field Physics and Department of Physics, National Central University, Taoyuan City 320317, Taiwan}
\fntext[fn_pensacola]{Present address: Pensacola State College, Pensacola, FL 32504, USA.}
\fntext[fn_slac]{Present address: Stanford Linear Accelerator Center, Menlo Park, CA 94025, USA.}
\fntext[fn_kagra]{Present address: Institute for Cosmic Ray Research, KAGRA Observatory, The University of Tokyo, Hida, Gifu 506-1205, Japan.}
\fntext[fn_gwu]{Present address: George Washington University, Washington, DC 20052, USA.}
\fntext[fn_nihon]{Present address: Nihon University, College of Science and Technology, Chiyoda-ku, Tokyo 101-8308, Japan.}
\fntext[fn_fnal]{Present address: Fermi National Accelerator Laboratory, Batavia, Illinois 60510, USA.}
\fntext[fn_aps]{Present address: APS, Argonne National Laboratory, Lemont, Illinois 60439, USA.}
\fntext[fn_tsinghua]{Present address: Department of Physics, Tsinghua University, Beijing 100084, China.}

\date{\today}

\begin{abstract}
	We report the $p+p$ and $p+d$ differential cross sections measured in the SeaQuest experiment
	for $J/\psi$ and $\psi\left(2S\right)$ production at \SI{120}{\GeV} beam energy covering the forward
	$x$-Feynman ($x_F$) range of $0.5 < x_F <0.9$. The measured cross sections are in good
	agreement with theoretical calculations based on the nonrelativistic QCD (NRQCD)
	using the long-distance matrix elements deduced from a recent global analysis
	of proton- and pion-induced charmonium production data. The $\sigma_{\psi\left(2S\right)} / \sigma_{J/\psi}$
	cross section ratios are found to increase as $x_F$ increases, indicating that the
	$q \bar{q}$ annihilation process has larger contributions in the $\psi\left(2S\right)$ production
	than the $J/\psi$ production. The $\sigma_{pd}/2\sigma_{pp}$ cross section ratios are
	observed to be significantly different for the Drell-Yan process and $J/\psi$ production,
	reflecting their different production mechanisms. We find that the $\sigma_{pd}/2\sigma_{pp}$
	ratios for $J/\psi$ production at the forward $x_F$ region are sensitive to the
	$\bar{d}/ \bar{u}$ flavor asymmetry of the proton sea, analogous to the Drell-Yan process.
	The transverse momentum ($p_T$) distributions for $J/\psi$ and $\psi\left(2S\right)$ production are also presented and
	compared with data collected at higher center-of-mass energies.
\end{abstract}

\maketitle

The SeaQuest experiment at Fermilab measures high-mass dimuons
produced in the interaction of a \SI{120}{\GeV} proton beam with various targets
including liquid hydrogen, liquid deuterium, and nuclear
targets~\cite{aidala2019}. Dimuons
originating from the Drell-Yan process~\cite{drell1970} and from the
decay of charmonium
states ($J/\psi$ and $\psi\left(2S\right)$) were collected simultaneously.
Results from SeaQuest on the $\sigma_{pd}/2\sigma_{pp}$
Drell-Yan cross section ratio,
which is sensitive
to the $\bar{d}/ \bar{u}$ flavor asymmetry in the proton, were reported
recently~\cite{dove2021,dove2023}. In this paper, we present results from
SeaQuest on the $J/\psi$ and $\psi\left(2S\right)$ charmonium production in $p+p$
and $p+d$ interactions.

Unlike the Drell-Yan process which primarily involves the
quark-antiquark annihilation through the electromagnetic interaction, charmonium
production proceeds via the strong interaction containing
both the quark-antiquark annihilation and the gluon-gluon
fusion processes. The simultaneous measurement of these two very different
processes provides complementary
information on the partonic structures of the nucleon. In particular,
the $\sigma_{pd}/2\sigma_{pp}$ ratio for charmonium
production is expected
to be sensitive to the ratio of the gluon distributions in the proton
and neutron, as well as the $\bar{d}/ \bar{u}$ ratio in the proton~\cite{peng1995}.

If quark-antiquark annihilation is an important subprocess for charmonium
production, then the $\sigma_{pd}/2\sigma_{pp}$ ratio would provide an
independent measurement of the $\bar{d}/ \bar{u}$ flavor asymmetry in the
proton~\cite{peng1995}, analogous to the Drell-Yan process.
On the other hand, the gluon-gluon fusion subprocess would allow the
$\sigma_{pd}/2\sigma_{pp}$ ratio to probe the relative gluon content
in the proton and neutron, providing a test of the charge symmetry (CS)
at the partonic level~\cite{piller1996}. The CS operation interchanges
the up and down quarks, and it also interchanges the proton and the neutron.
Since the gluon is an iso-scalar particle, CS requires that the gluon distributions
in the proton and neutron are identical.
Violation of CS is predicted at both the hadronic~\cite{stephenson2003,opper2003}
and the partonic~\cite{londergan2010} levels.
A measurement of the gluon contents of the proton and neutron could test CS at
the partonic level~\cite{piller1996,zhu2008,lansberg2012}.
It is also interesting to compare the production mechanisms for $J/\psi$ versus $\psi\left(2S\right)$.

While proton-induced charmonium production is often dominated
by the gluon-gluon fusion process~\cite{vogt1999}, the
quark-antiquark annihilation process could also contribute significantly.
The relative importance of these two
processes depends on the beam energy and on the
$x$-Feynman ($x_F$) (see \cref{eq:eq1}) of the charmonium~\cite{peng1995},
and can be calculated using various production models:
color evaporation model (CEM)~\cite{einhorn1975,fritzsch1977,halzen1977}, color singlet model (CSM)~\cite{chang1980},
and nonrelativistic QCD (NRQCD)~\cite{bodwin1995}.
These various models predict different  relative importance of the different subprocesses~\cite{vogt2000}.
In this paper, we mainly focus on the comparison with NRQCD, with the comparison with CEM shown in the Supplemental Material.

The NA51 collaboration reported a measurement of the $p+p$ and $p+d$
cross sections for charmonium production at \SI{450}{\GeV} at a single value of
$x_F \approx 0$~\cite{abreu1998}. The SeaQuest measurement covers the
broader kinematic range of $0.5 < x_F < 0.9$ at the lower beam energy of
\SI{120}{\GeV}. These two measurements can provide complementary information.

The SeaQuest experiment was performed using the \SI{120}{\GeV} proton beam from the
Fermilab Main Injector. The SeaQuest dimuon spectrometer was designed for
detecting high-mass dimuon pairs produced in the interaction of a proton
with various targets. Details of the SeaQuest spectrometer can be found
elsewhere~\cite{aidala2019,dove2021,dove2023}. A primary proton beam containing
up to $6 \times 10^{12}$ protons in a 4-second long beam spill every minute
was incident upon one of three identical \SI{50.8}{\cm} long cylindrical
stainless steel target flasks or solid nuclear targets.
The targets alternated between liquid hydrogen, liquid deuterium, solid nuclear targets,
and the empty flask target. A Cherenkov counter
was placed in the beam to record the instantaneous proton intensity for
each 1-ns long RF bucket at a \SI{53}{\MHz} repetition rate.

\begin{figure}[tb]
	\includegraphics*[width=\linewidth]{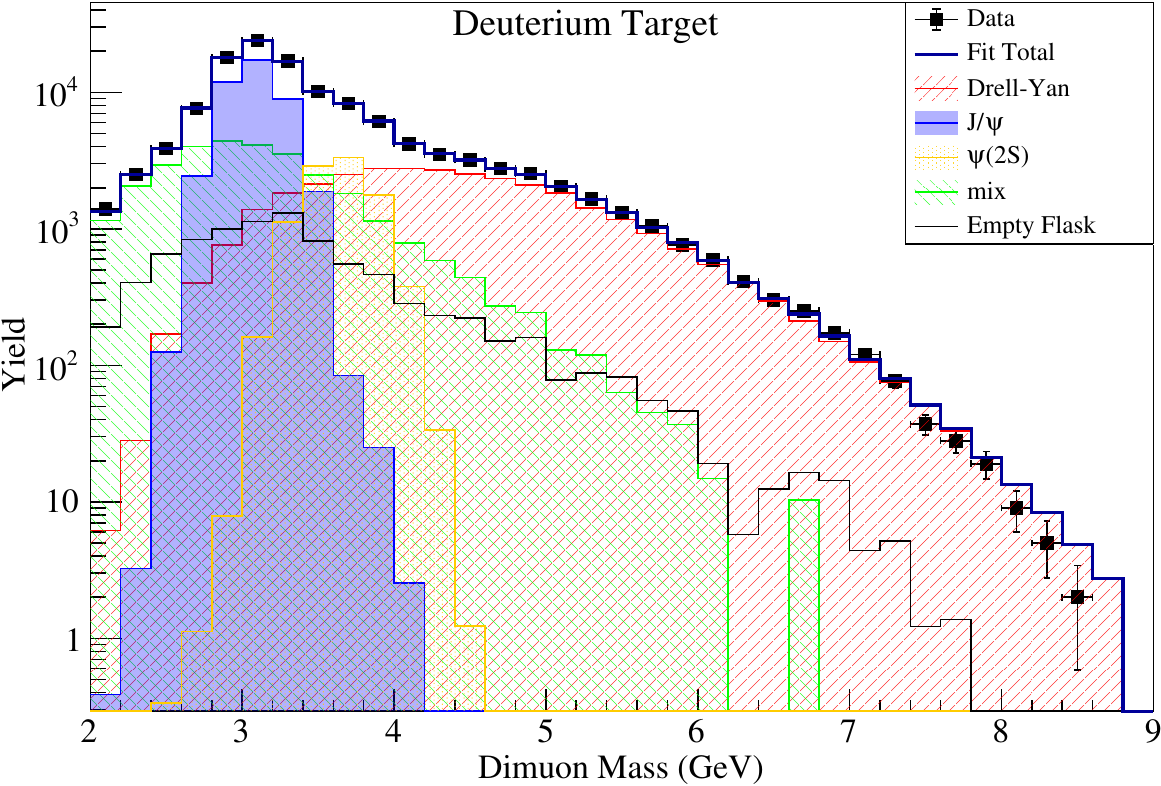}
	\caption{Dimuon mass distribution for events collected
		on a liquid deuterium target for the second data set.
		The data points (solid squares) are
		compared with a fit (solid blue line) consisting of
		various components (see text).}
	\label{fig:LD2_Mass}
\end{figure}

The SeaQuest spectrometer consists of two dipole magnets and
four detector stations equipped with hodoscopes and tracking chambers.
A solid iron magnet downstream of the target focuses the dimuons and
also serves as a beam-dump and a hadron absorber. An open magnet further
downstream measures the muon momentum. The dimuon trigger
requires a quadruple hodoscope coincidence with a pattern consistent with
a muon pair originating from the target. Various diagnostic triggers
are also implemented. In particular, the ``single-muon" trigger
is used to evaluate the accidental dimuon
background, and the ``random" trigger samples the
detector response throughout the data-taking periods.

\begin{figure}[tb]
	\includegraphics*[width=\linewidth]{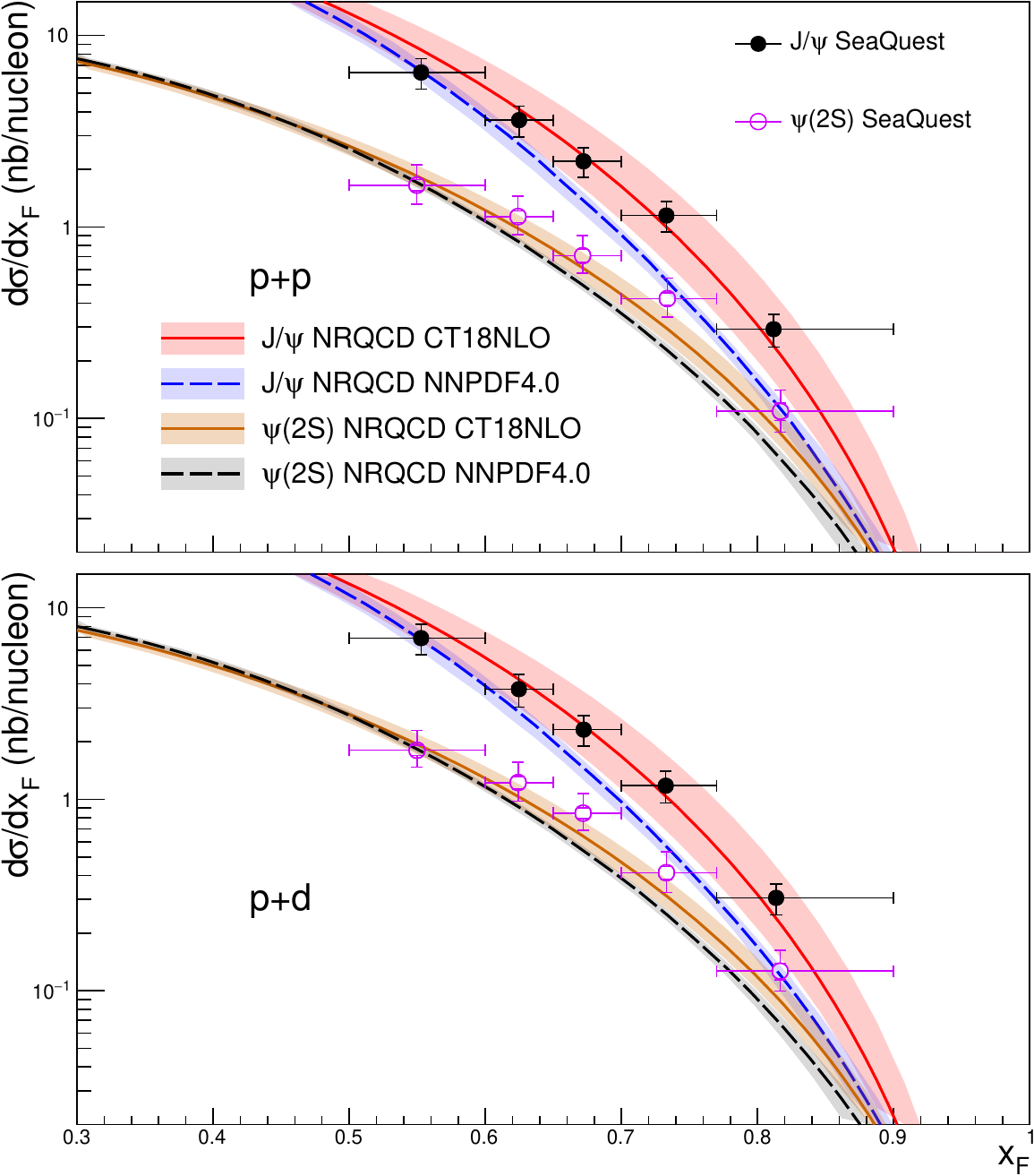}
	\caption{The differential cross
		section per nucleon $d\sigma / d x_F$ for $J/\psi$ and $\psi\left(2S\right)$
		production in $p+p$ and $p+d$ interactions at \SI{120}{\GeV}, integrated over $p_T$.
		The error bars represent the total uncertainties.
		The curves correspond to NRQCD calculation~\cite{beneke1996} using the LDMEs obtained
		in~\cite{chang2023} and the nucleon PDFs from CT18~\cite{hou2021} and
		NNPDF4.0~\cite{ball2022a}. The error bands indicate 68\% confidence level from the PDFs.}
	\label{fig:cs_xF}
\end{figure}

\begin{figure}[tb]
	\includegraphics[width=\linewidth]{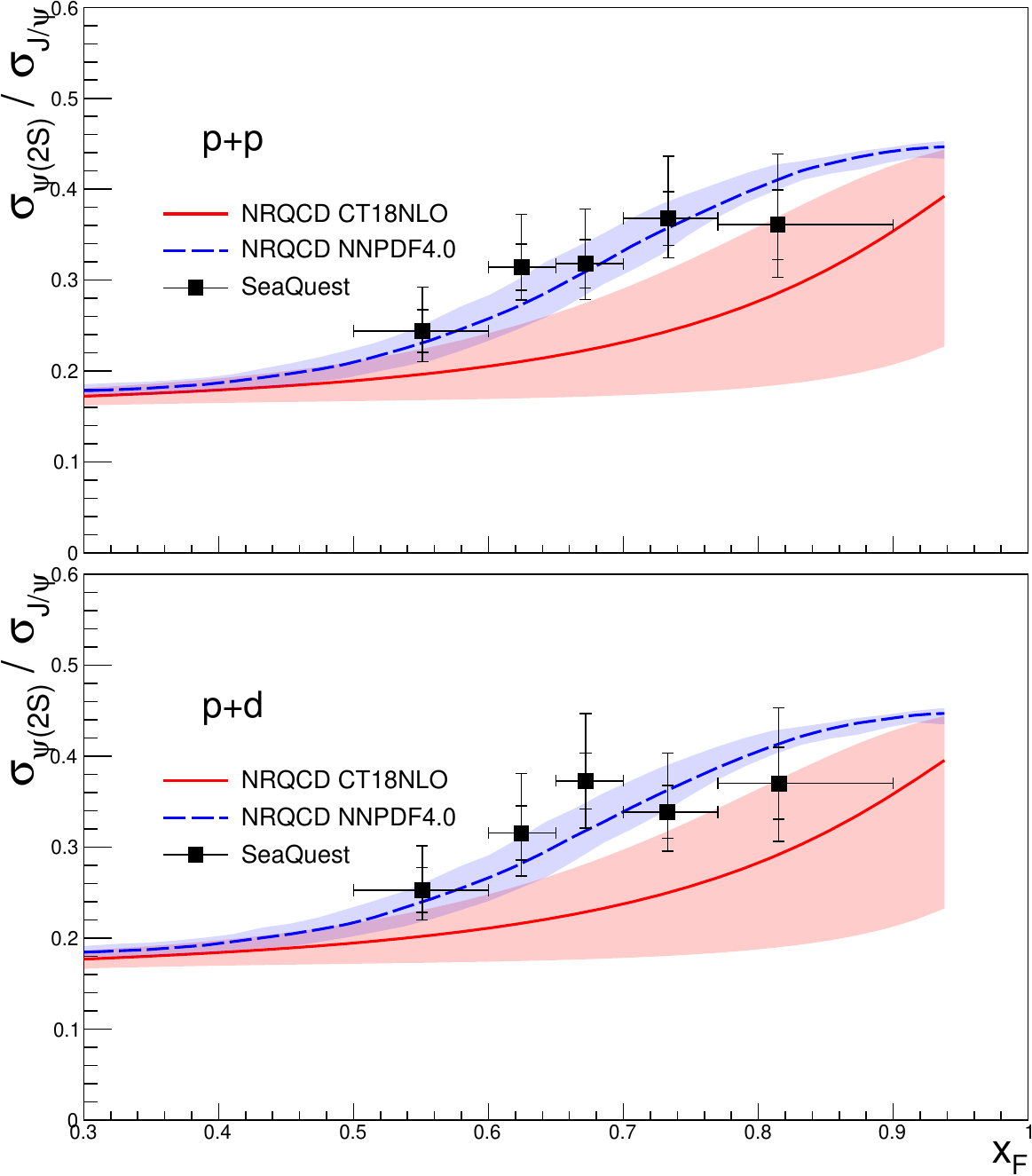}
	\caption{The ratio of $\sigma_{\psi\left(2S\right)}/\sigma_{J/\psi}$ in $p+p$ and $p+d$
		interactions at \SI{120}{\GeV}.
		The inner (outer) error bars represent the statistical (total) uncertainties.
		The curves correspond to NRQCD calculation
		using CT18 and NNPDF4.0. The error bands indicate 68\% confidence level from the PDFs.}
	\label{fig:psi_ratio}
\end{figure}

The SeaQuest data are separated into two sets, each containing
roughly half of the total data sample. The first part includes data
taken between June 2014 and July 2015, and the second part covers the
remaining period up to July 2017.
Results on the analysis of the Drell-Yan events from the first data set
have already been reported~\cite{dove2021,dove2023}. In this paper,
we have analyzed the full SeaQuest data sets.
Since the trigger conditions and the detector configuration for the two
data sets are not identical, the analysis was performed
separately for each data set. Results obtained from the two data sets
are first compared to verify their consistency, and then combined for
the final results.

\begin{figure}[tb]
	\includegraphics[width=\linewidth]{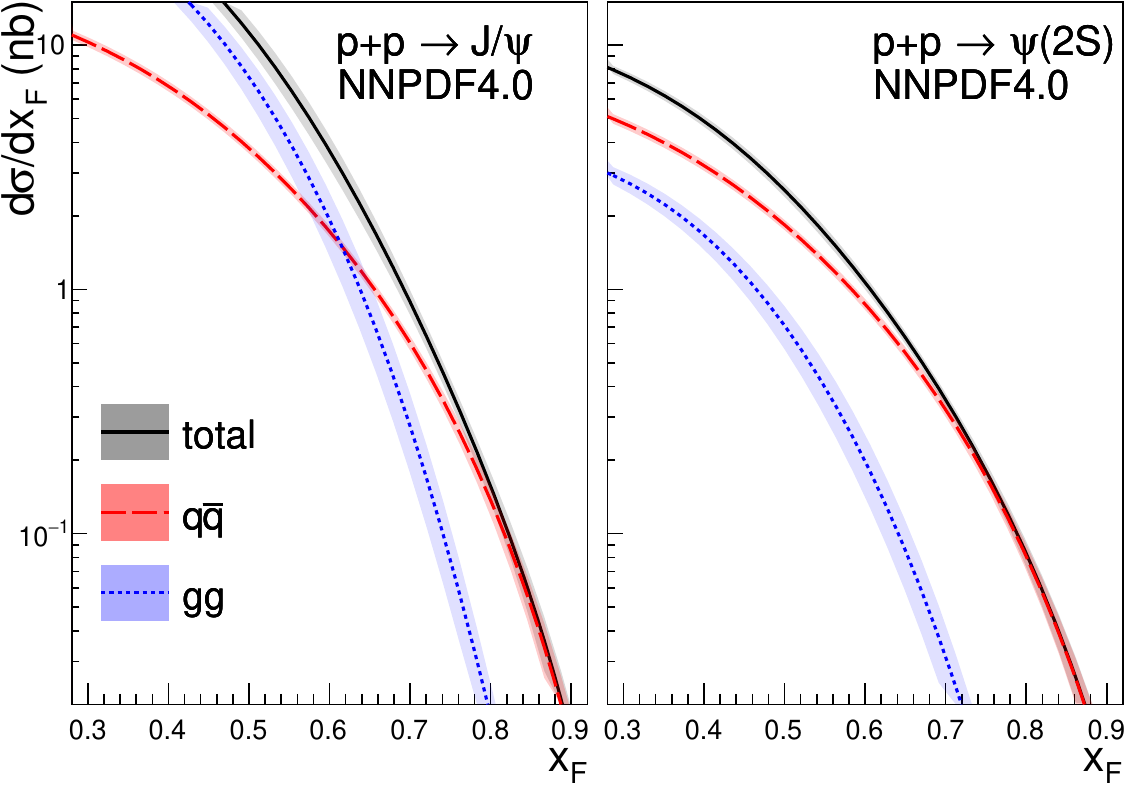}
	\caption{The individual contribution from the $q\bar{q}$ (red dash) and $gg$
		(blue dotted) process to the production of $J/\psi$ (left) and $\psi\left(2S\right)$ (right)
		calculated in NRQCD using NNPDF4.0 PDFs.}
	\label{fig:NRQCD_process}
\end{figure}

\begin{figure}[tb]
	\includegraphics*[width=\linewidth]{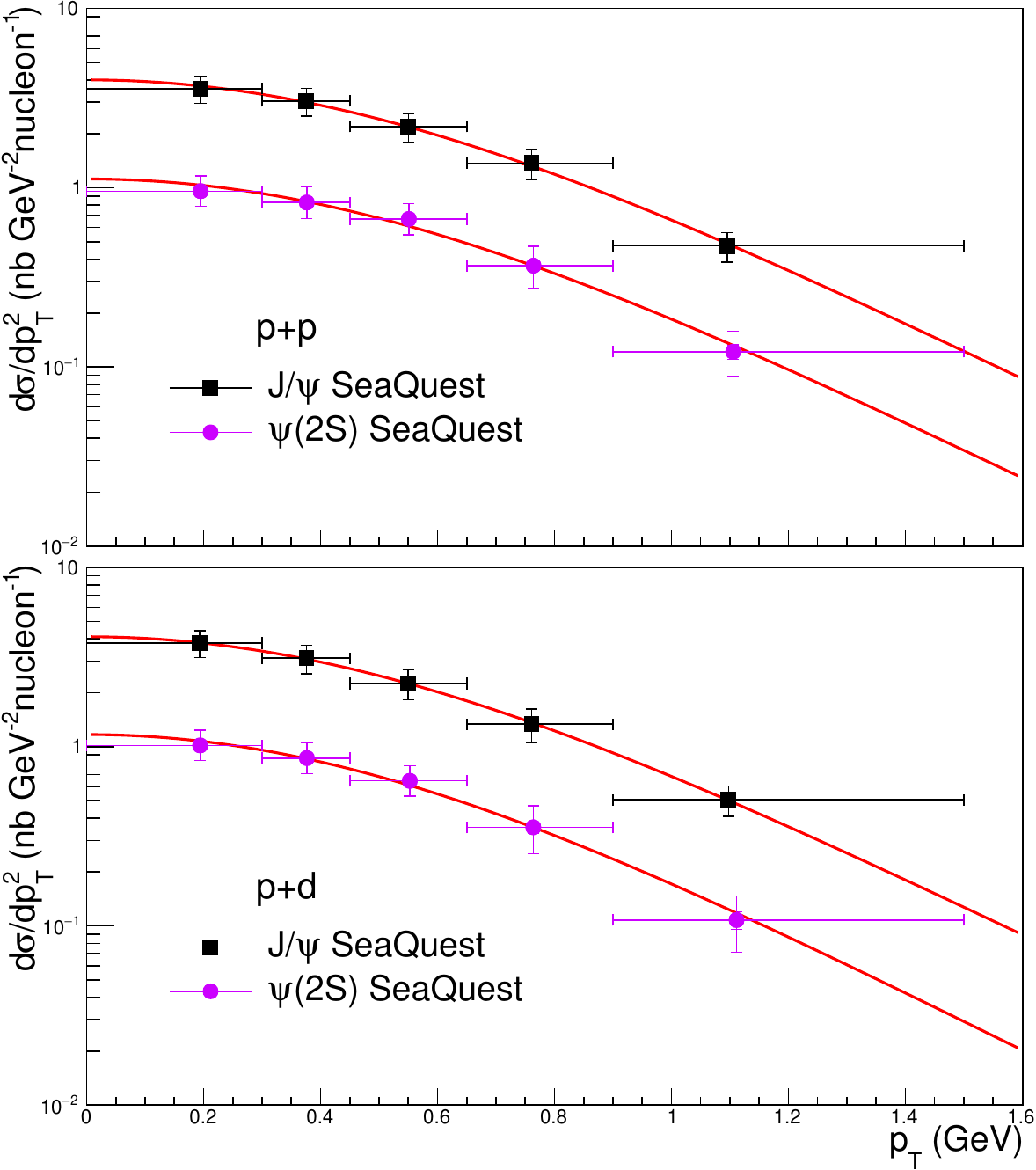}
	\caption{The differential cross section per nucleon
		$d\sigma / d p^2_T$, integrated over $0.5<x_{F}<0.9$, for $J/\psi$ and $\psi\left(2S\right)$
		production in $p+p$ and $p+d$ interaction at \SI{120}{\GeV}.
		The error bars represent the total uncertainties.
		The curves correspond to fits using the Kaplan form described in the
		text.}
	\label{fig:cs_pT}
\end{figure}

Details of the data analysis procedure can be found
in Refs.~\cite{dove2021,dove2023}.
Candidate muon tracks reconstructed in the drift chambers are extrapolated
to the target region. Only dimuon events consistent with originating
from the target are selected. The target position is then used to
refine the parameters of each muon pair. The resulting
RMS mass resolution for $J/\psi$ is $\approx$\SI{200}{\MeV},
dominated by the finite target length and the
multiple scattering of muons in the iron magnet.

\Cref{fig:LD2_Mass} shows the dimuon mass spectrum for $p+d$
data collected in the second data set. A comparison with the
mass spectrum obtained for the first data set,
reported in Ref.~\cite{dove2023}, shows
good agreement with
some small differences attributed to the minor changes in
trigger conditions and spectrometer settings.

To extract the yields of $J/\psi$ and
$\psi\left(2S\right)$, the dimuon mass spectrum is fitted by
including several components. First,
data collected with the empty target flask
are analyzed to determine the background originating
from sources other than the liquid.
Second, a GEANT4~\cite{agostinelli2003,allison2006,allison2016} based Monte Carlo (MC) simulation
is performed to obtain the expected line shapes of the $J/\psi$
and $\psi\left(2S\right)$ resonances. The MC
dimuon events are then embedded with additional hits in the detectors
using data collected with the ``random" trigger,
which randomly samples the spectrometer response to background hits.
This procedure accounts for the spectrometer response to background hits.
Third, dimuons from the Drell-Yan process are simulated using
a next-to-leading order
calculation~\cite{catani2009} with the CT14 parton distribution functions
(PDFs)~\cite{dulat2016}, as described in an earlier publication~\cite{dove2023}.
The embedding procedure is also applied to the Drell-Yan MC data.
Finally, the accidental dimuon background, caused by two independent
interactions within the same RF bucket, is simulated
by forming a random combination of data collected with the
``single-muon" trigger, as discussed in detail in Ref.~\cite{dove2023},
labeled as ``mix'' in \cref{fig:LD2_Mass}.
Other mixing methods~\cite{pate2023} have also been studied and included in the systematic uncertainties.
These embedded MC events are then analyzed by applying cuts identical
to those for the real data.

A fit to the $p+d$ dimuon data,
allowing the normalizations of
the various components except the empty flask data to vary, is
shown in \cref{fig:LD2_Mass}.
The empty flask data are normalized according to their relative luminosity.
The data are well described as the sum of various
components. The adequacy of this approach is further validated by the excellent
agreement between this method~\cite{dove2023} and an independent
intensity-extrapolation method~\cite{dove2021} for the extracted
$\sigma_{pd}/2\sigma_{pp}$ Drell-Yan cross section ratios.

\begin{table*}
	\centering
	\caption{The differential cross sections per nucleon, $d\sigma/dx_F$
		(in \unit{\nano\barn}), for $J/\psi$ and $\psi\left(2S\right)$ production in $p+p$ and $p+d$
		collisions at \SI{120}{\GeV} for different $x_F$ bins.
		The statistical uncertainties followed by systematic uncertainties are also shown.}
	\label{tab:xfcros}
	\begin{adjustbox}{max width=\linewidth}
		\renewcommand{\arraystretch}{1.5}
\begin{tabular}{cccc|cccc}
\hline
\multicolumn{4}{c|}{$p+p$}                                           & \multicolumn{4}{c}{$p+d$}                                            \\ \hline
\rule{0pt}{2.6ex}\rule[-1.5ex]{0pt}{0pt}
$\expval{x_F}_{J/\psi}$ &
  $\eval{d\sigma/dx_F}_{J/\psi}$ &
  $\expval{x_F}_{\psi\left(2S\right)}$ &
  $\eval{d\sigma/dx_F}_{\psi\left(2S\right)}$ &
  $\expval{x_F}_{J/\psi}$ &
  $\eval{d\sigma/dx_F}_{J/\psi}$ &
  $\expval{x_F}_{\psi\left(2S\right)}$ &
  $\eval{d\sigma/dx_F}_{\psi\left(2S\right)}$ \\ \hline
0.553 & $6.411\pm0.246\pm1.130$ & 0.550 & $1.654\pm0.112^{+0.451}_{-0.319}$ & 0.553 & $6.944\pm0.275\pm1.224$ & 0.550 & $1.802\pm0.112^{+0.468}_{-0.315}$ \\
0.625 & $3.618\pm0.145\pm0.647$ & 0.624 & $1.134\pm0.079^{+0.302}_{-0.209}$ & 0.625 & $3.758\pm0.166\pm0.706$ & 0.624 & $1.222\pm0.088^{+0.329}_{-0.230}$ \\
0.672 & $2.204\pm0.082\pm0.383$ & 0.671 & $0.709\pm0.055^{+0.184}_{-0.124}$ & 0.672 & $2.309\pm0.087\pm0.408$ & 0.672 & $0.846\pm0.055^{+0.220}_{-0.148}$ \\
0.733 & $1.149\pm0.037\pm0.205$ & 0.734 & $0.423\pm0.031^{+0.113}_{-0.079}$ & 0.733 & $1.177\pm0.039\pm0.217$ & 0.733 & $0.413\pm0.032^{+0.114}_{-0.082}$ \\
0.812 & $0.293\pm0.011\pm0.056$ & 0.817 & $0.109\pm0.012^{+0.030}_{-0.021}$ & 0.814 & $0.305\pm0.013\pm0.055$ & 0.817 & $0.127\pm0.013^{+0.034}_{-0.024}$ \\ \hline
\end{tabular} 	\end{adjustbox}
\end{table*}

To obtain the charmonium differential cross sections,
the data were split into bins of $x_F$ and $p_T$ and the dimuon mass
spectrum for each bin is fitted with the procedure described earlier
to extract the $J/\psi$ and $\psi\left(2S\right)$ yields.
We note the following definition of $x_F$~\cite{dove2021}:
\begin{equation}
	x_F = \frac{2p_L}{\sqrt{s}\left(1-M^2/s\right)},
	\label{eq:eq1}
\end{equation}
where $p_L$ is the longitudinal momentum of the dimuon in the hadron-hadron center of mass frame.
$M$ and $\sqrt{s}$ are the dimuon mass and the hadron-hadron total
energy, respectively.
The charmonium production cross section is obtained as
\begin{equation}
	d\sigma = \frac{dY}{B\cdot \mathrm{Acc} \cdot \mathrm{Eff} \cdot \mathrm{Lum}},
	\label{eq:eq2}
\end{equation}
where the yield $dY$ is the number of $J/\psi$ or $\psi\left(2S\right)$ events
for each $x_F$ or $p_T$ bin, $\mathrm{Acc}$ the spectrometer acceptance,
$\mathrm{Eff}$ the efficiency for  analysis cuts, $\mathrm{Lum}$ the effective
luminosity including the data-acquisition deadtime, and $B$
the branching ratio for decaying into a muon pair. We use
$B\left(J/\psi\to\mu^+\mu^-\right)=\left(5.961\pm0.033\right)\%$
and $B\left(\psi\left(2S\right)\to\mu^+\mu^-\right) = \left(8.0\pm0.6\right)\times10^{-3}$~\cite{workman2022}.

\begin{table*}
	\centering
	\caption{The differential cross sections per nucleon, $d\sigma / dp^2_T$
		(in \unit{\nano\barn\per\GeV\squared}),
		for charmonium production in $p+p$ and $p+d$
		collisions at \SI{120}{\GeV} for different $p_T$ bins.
		The statistical uncertainties followed by systematic uncertainties are also shown.}
	\label{tab:ptcros}
	\begin{adjustbox}{max width=\linewidth}
		\renewcommand{\arraystretch}{1.5}
\begin{tabular}{cccc|cccc}
\hline
\multicolumn{4}{c|}{$p+p$}                                        & \multicolumn{4}{c}{$p+d$}                                         \\ \hline
\rule{0pt}{2.6ex}\rule[-1.5ex]{0pt}{0pt}
$\expval{p_T}_{J/\psi}$ &
  $\eval{d\sigma/dp_T^2}_{J/\psi}$ &
  $\expval{p_T}_{\psi\left(2S\right)}$ &
  $\eval{d\sigma/dp_T^2}_{\psi\left(2S\right)}$ &
  $\expval{p_T}_{J/\psi}$ &
  $\eval{d\sigma/dp_T^2}_{J/\psi}$ &
  $\expval{p_T}_{\psi\left(2S\right)}$ &
  $\eval{d\sigma/dp_T^2}_{\psi\left(2S\right)}$ \\ \hline
0.195 & $3.570\pm0.134\pm0.605$ & 0.195 & $0.957\pm0.059^{+0.200}_{-0.160}$ & 0.193 & $3.789\pm0.137\pm0.634$ & 0.194 & $1.017\pm0.055^{+0.211}_{-0.169}$ \\
0.376 & $3.045\pm0.114\pm0.519$ & 0.376 & $0.827\pm0.043^{+0.181}_{-0.149}$ & 0.376 & $3.119\pm0.115\pm0.558$ & 0.377 & $0.864\pm0.046^{+0.185}_{-0.150}$ \\
0.550 & $2.196\pm0.070\pm0.392$ & 0.551 & $0.669\pm0.031^{+0.145}_{-0.119}$ & 0.550 & $2.251\pm0.071\pm0.421$ & 0.553 & $0.645\pm0.030^{+0.137}_{-0.111}$ \\
0.761 & $1.373\pm0.052\pm0.261$ & 0.764 & $0.367\pm0.023^{+0.102}_{-0.091}$ & 0.761 & $1.337\pm0.056\pm0.277$ & 0.764 & $0.355\pm0.025^{+0.109}_{-0.099}$ \\
1.095 & $0.473\pm0.021\pm0.086$ & 1.106 & $0.122\pm0.011^{+0.035}_{-0.031}$ & 1.098 & $0.506\pm0.022\pm0.096$ & 1.111 & $0.108\pm0.012^{+0.037}_{-0.035}$ \\ \hline
\multicolumn{2}{c}{
\rule{0pt}{2.6ex}\rule[-1.5ex]{0pt}{0pt}
$\expval{p_T^2}=0.714\pm0.021\pm0.040$} &
  \multicolumn{2}{c|}{$\expval{p_T^2}=0.714\pm0.036\pm0.068$} &
  \multicolumn{2}{c}{$\expval{p_T^2}=0.717\pm0.022\pm0.046$} &
  \multicolumn{2}{c}{$\expval{p_T^2}=0.663\pm0.035\pm0.067$} \\ \hline
\end{tabular} 	\end{adjustbox}
\end{table*}

\begin{figure}[tb]
	\includegraphics[width=\linewidth]{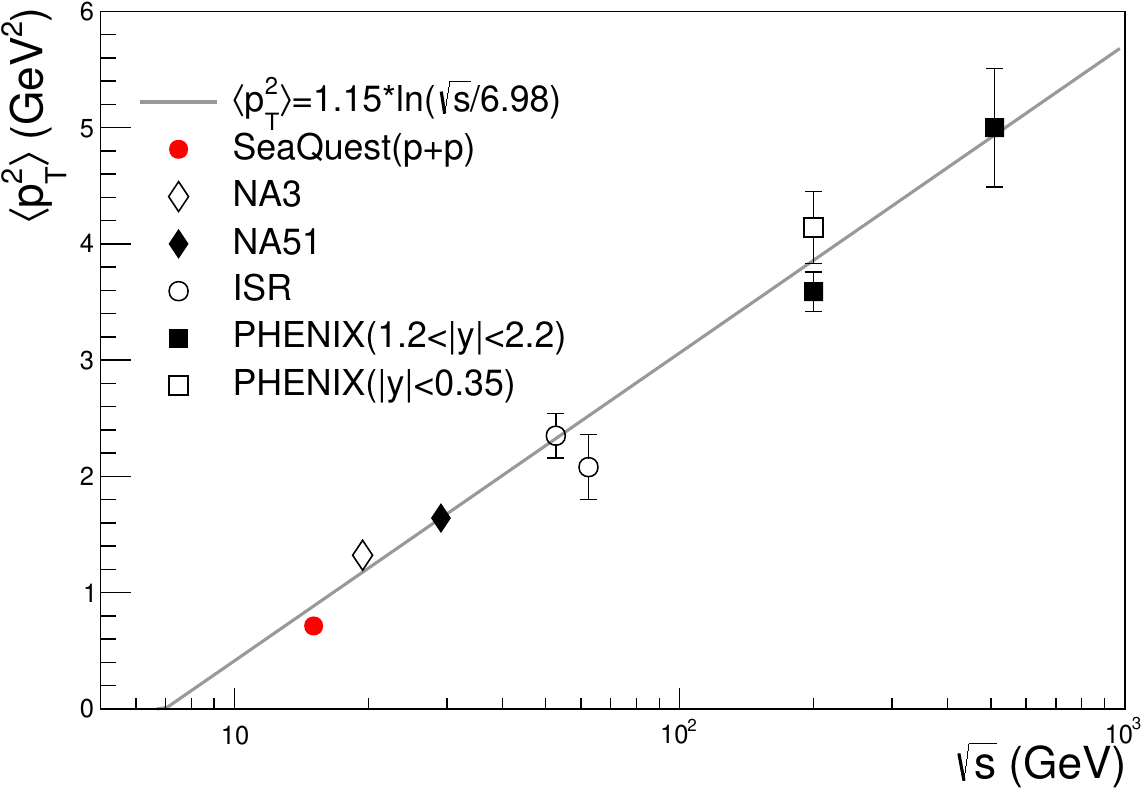}
	\caption{The extracted $\expval{p_T^2}$ for $p+p\to J/\psi$ from SeaQuest (solid red circle) compared
		to other experiments~\cite{badier1983,drapier1998,clark1978,acharya2020} at different $\sqrt{s}$.
		The $\expval{p_T^2}$ increases logarithmically versus $\sqrt{s}$,
		as illustrated by the fit (gray line) to the data.}
	\label{fig:pT_s}
\end{figure}

The $x_F$ dependence of the $J/\psi$ and
$\psi\left(2S\right)$ production cross sections in $p+p$ and $p+d$ collisions
is shown in
\cref{fig:cs_xF} and listed
in \cref{tab:xfcros}. 
In this and the subsequent figures, the horizontal error bars represent the bin width,
and the data points are positioned on the ordinate at the mean value for the events in the bin.
The $d\sigma/ d x_F$ differential cross sections are
obtained with an acceptance calculation using a $p_T$ distribution
which best fits the data. The systematic uncertainties
include an overall normalization uncertainty, common to both $p+p$ and $p+d$ cross sections.
Other uncertainties which are largely independent of data set are the relative normalization of the flask data,
the event mixing procedure ($\approx 7.2\%$), the trigger efficiency ($\approx 11\%$),
reconstruction efficiency ($\approx 15\%$) and the trigger roadset dependence ($\approx 2\%$).
A second set of uncertainties correlated between data sets are the $J/\psi$ and $\psi\left(2S\right)$ polarization ($\approx 5.5\%$) 
and the uncertainty in the beam normalization ($\approx 10\%$).
More discussion on the systematic uncertainties can be found in Ref.~\cite{leung2024}.

\begin{figure}[tb]
	\includegraphics*[width=\linewidth]{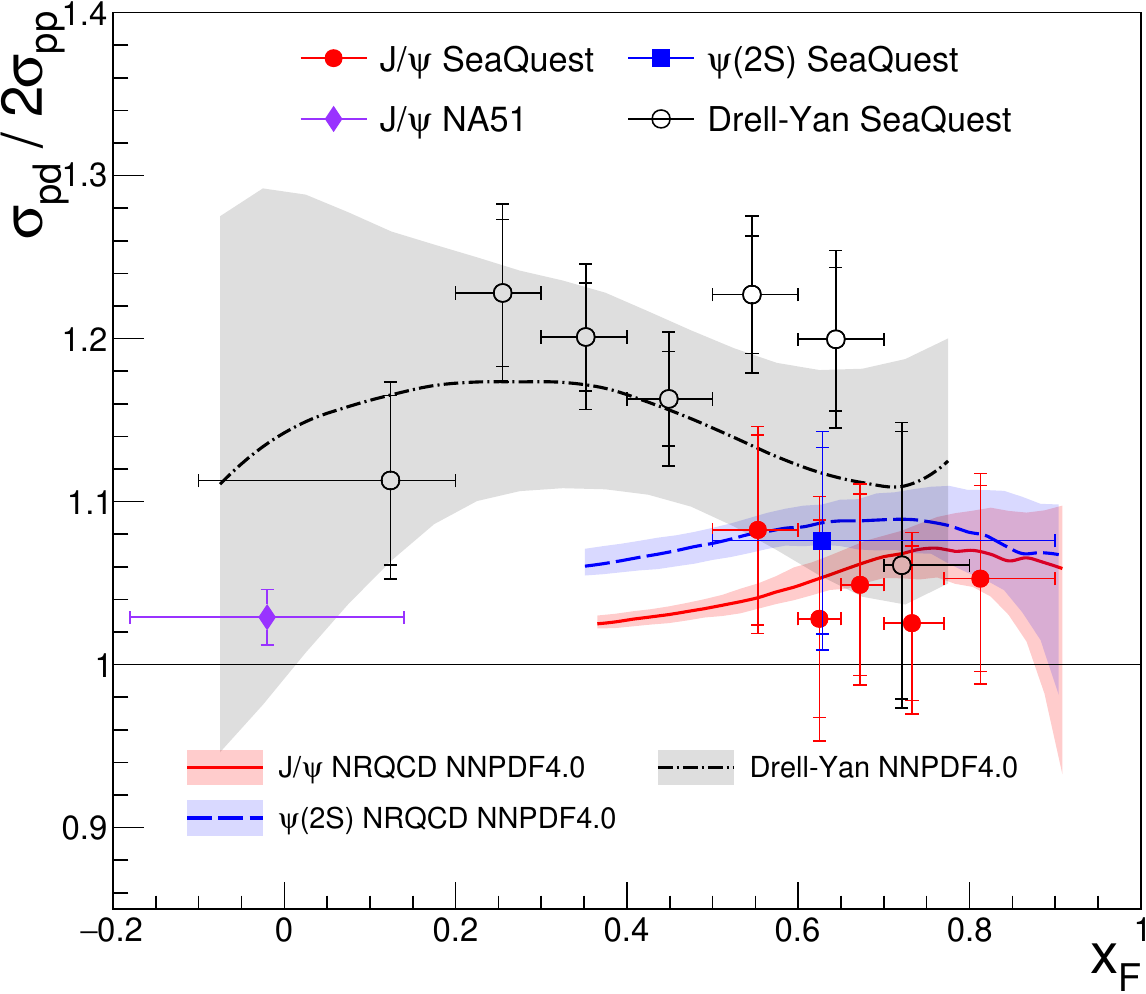}
	\caption{The $\sigma_{pd}/2\sigma_{pp}$ cross section ratios for $J/\psi$ (red circle)
		and $\psi\left(2S\right)$ (blue square) versus $x_F$ from SeaQuest.
		The inner (outer) error bars represent the statistical (total) uncertainties.
		For comparison, the ratio for $J/\psi$ measured by NA51~\cite{abreu1998}
		and the ratios for the Drell-Yan cross sections by SeaQuest~\cite{dove2023} are
		also shown. The solid (dashed) curve is the calculation for the $J/\psi$  ($\psi\left(2S\right)$) cross section
		ratios at \SI{120}{\GeV} in the NRQCD model using the proton PDFs
		from NNPDF4.0~\cite{ball2022a}.
		The dot-dashed curve represents the NLO Drell-Yan cross section ratios
		calculated with the same PDFs.
		The error bands indicate 68\% confidence level from the PDFs.}
	\label{fig:pdpp}
\end{figure}

The $d\sigma/dx_F$ distributions of charmonium production are compared
with theoretical calculations in \cref{fig:cs_xF}. The calculations were performed
using the non-relativistic QCD (NRQCD)~\cite{bodwin1995,maltoni2006} approach,
which is based on
the factorization of the heavy-quark $Q \bar{Q}$ pair production
and its subsequent hadronization. The $Q \bar{Q}$ production includes
the subprocesses of gluon-gluon fusion, quark-antiquark annihilation,
and quark-gluon interaction. The hadronizations
into quarkonium bound states are described by a set of
long-distance matrix elements (LDMEs), assumed to be universal
and fixed by the experimental data~\cite{beneke1996,maltoni2006}.
The LDMEs are taken from a recent global fit to fixed-target proton
and pion induced $J/\psi$ and $\psi\left(2S\right)$ production data performed with the SMRS pion and CT14 proton PDFs at charm mass $m_c=\SI{1.5}{\GeV}$
in Ref.~\cite{chang2023}, which give the best overall $\chi^2$ in their analysis.
The estimated $J/\psi$ cross section also includes the feed-down from
hadronic decays of $\psi\left(2S\right)$ and radiative decays of the three $\chi_{cJ}$ states as described in Ref.~\cite{chang2023}.
\Cref{fig:cs_xF} shows that the $d\sigma / d x_F$ data for $p+p$ and $p+d$ are very well
described by the NRQCD calculation~\cite{beneke1996,maltoni2006} using CT18~\cite{hou2021},
including the overall normalization, which is fixed by the LDMEs.
The extracted cross sections are also compared to the color evaporation model (CEM)~\cite{nason1988,nason1989,mangano1993,schuler1996,nelson2013}
in \cref{fig:cs_xF_CEM}. 
In the CEM framework, the hadronization probability is independent of the underlying sub-process,
and it is typically obtained from fitting to data.
As shown in \cref{fig:cs_xF_CEM} of the Supplemental Material, the measured $J/\psi$ $x_F$ distributions are also in good agreement with CEM calculations,
but the CEM calculations tend to underestimate the $\psi\left(2S\right)$ cross section at large $x_F$.

The ratio of $\sigma_{\psi\left(2S\right)}/\sigma_{J/\psi}$ as a function of $x_F$ is
shown in \cref{fig:psi_ratio}. The measured ratio is found to increase
as $x_F$ increases, suggesting a broader $x_F$ distribution for $\psi\left(2S\right)$
production than for $J/\psi$ production. This behavior is well described
by the NRQCD calculations as shown in \cref{fig:psi_ratio}. Since the valence
quark in the proton has a much broader $x$ distribution than the gluon, one
expects the $q \bar q$ annihilation process would give a broader $x_F$
distribution than the gluon-gluon fusion process. Therefore, the broader
$x_F$ distribution for $\psi\left(2S\right)$ production is attributed to the
increasing importance of the $q \bar q$ annihilation process for $\psi\left(2S\right)$
production. This implies that the $\psi\left(2S\right)$ production
is more analogous to the Drell-Yan process, which is dominated by the $q\bar{q}$
annihilation process.
\Cref{fig:NRQCD_process} shows the individual contribution from the $q\bar{q}$ annihilation
and gluon fusion processes to the $J/\psi$ and $\psi\left(2S\right)$ production in $p+p$ collision calculated
in NRQCD using NNPDF4.0 PDFs.
At $x_F\leq 0.6$, the gluon fusion process is more important than $q\bar{q}$ annihilation for $J/\psi$ production.
In contrast, the $\psi\left(2S\right)$ production is dominated by the $q\bar{q}$ annihilation process for the entire $x_F$ range.
Similar behavior was also observed for pion-induced $J/\psi$ and $\psi\left(2S\right)$ production data~\cite{heinrich1991}.

It should be noted that the relative importance of these subprocesses remains uncertain
and depends on the production model used~\cite{vogt2000}.
Unlike NRQCD, the fragmentation probability in CEM is independent of the underlying sub-process,
and only depends on the final charmonium state. 
Hence, CEM would suggest the $x_F$ distribution to be identical for $J/\psi$ and $\psi\left(2S\right)$,
except for the relative fraction of $\psi\left(2S\right)$ production.
Therefore CEM would predict the $\sigma_{\psi\left(2S\right)}/\sigma_{J/\psi}$ ratio to be independent of $x_F$,
which qualitatively disagrees with the data shown in \cref{fig:psi_ratio}.

The $p_T$ dependence of the $J/\psi$ and $\psi\left(2S\right)$
cross sections is shown in \cref{fig:cs_pT} and listed in \cref{tab:ptcros}
for $p+p$ and $p+d$. The $d \sigma / d p_T^2$ differential
cross sections are obtained by using the $x_F$ distribution obtained
from NRQCD to evaluate the spectrometer acceptance for $J/\psi$
and $\psi\left(2S\right)$. These $p_T$ distributions are fitted with
the Kaplan parameterization $d\sigma/dp_T^2 = c \left(1+p_T^2/p_0^2\right)^{-6}$~\cite{kaplan1978}
and the results of the fits are shown in \cref{fig:cs_pT}.
The $\expval{p_T}$ and $\expval{p_T^2}$ can be expressed as
\begin{equation}
	\expval{p_T}=\frac{35\pi p_0}{256},\quad
	\expval{p_T^2}=\frac{p^2_0}{4}.
\end{equation}
And the values of $\expval{p^2_T}$ are also listed in \cref{tab:ptcros},
showing very
similar values for $p+p$ and $p+d$, as well as for $J/\psi$
and $\psi\left(2S\right)$.
While the $p_T$ distributions cannot be reliably calculated for $p_T\ll M$ with fixed-order perturbative calculations, 
nonetheless, they could be calculated within the NRQCD framework by including the soft-gluon resummation,
as outlined in Ref.~\cite{sun2013}.
It would be interesting to compare our results on the $x_F$ and $p_T$ dependence 
with NRQCD calculations including soft-gluon resummation.

The extracted $\expval{p_T^2}$ for $p+p\to J/\psi$ is compared with
results from NA3~\cite{badier1983}, NA51~\cite{drapier1998}, ISR~\cite{clark1978}
and PHENIX~\cite{acharya2020} in \cref{fig:pT_s}. The $\expval{p_T^2}$
increases logarithmically as $\sqrt{s}$ increases over a wide range of energies.
A linear fit versus the log of the center-of-mass energy, adapted from Ref.~\cite{acharya2020},
\begin{equation}
	\expval{p_T^2} = a \ln\left(\sqrt{s}/b\right),
\end{equation}
with
\begin{equation}
	a=\SI{1.150\pm0.043}{\GeV\squared},\, b=\SI{6.98\pm0.37}{\GeV},
\end{equation}
describes the general trend. Some variation is expected due to the differing
rapidity range of the measurements, as shown in previous fixed-target $J/\psi$
production measurements~\cite{biino1987}.

The $\sigma_{pd}/2\sigma_{pp}$ $J/\psi$ cross section ratios versus $x_F$
are shown in \cref{fig:pdpp}.
As a result of the identical target geometry
of the two liquid targets and the frequent interchange between the targets,
most of the systematic uncertainties largely cancel in the
cross section ratio. The remaining systematic uncertainties shown in \cref{fig:pdpp}
have dominant contributions from the uncertainties
associated with the mass-fitting procedure.
Also shown in \cref{fig:pdpp} is the $J/\psi$ cross section
ratio at $x_F \approx 0$ measured by the NA51 collaboration~\cite{abreu1998} with
the \SI{450}{\GeV} proton beam.
The average ratio for the $J/\psi$ production across the SeaQuest measured region is $\approx 1.055\pm 0.033\pm0.025$.
Both the SeaQuest and the NA51 data show that the $\sigma_{pd}/2\sigma_{pp}$
ratios for $J/\psi$ production are greater than unity with $\approx2\sigma$ significance.

The $\sigma_{pd}/2\sigma_{pp}$ cross section ratio for $J/\psi$ and $\psi\left(2S\right)$ production is also compared with the
Drell-Yan process~\cite{dove2023} in \cref{fig:pdpp}.
The difference between the Drell-Yan and the charmonium
cross section ratios in \cref{fig:pdpp} clearly reflects the different underlying
mechanisms in these two processes. The Drell-Yan process, dominated by the
$q \bar{q}$  annihilation subprocess, leads to the expectation
that the cross section ratio is
approximately $\left(1+\bar{d}(x_2)/\bar{u}(x_2)\right)/2$ at forward $x_F$~\cite{dove2023}, where $x_2$ is the momentum
fraction of the parton in the target proton.
The measured range of $0.5<x_F<0.9$ corresponds to $0.048<x_2<0.078$ for $J/\psi$ production,
which covers a region of $x_2$ smaller than that covered by the Drell-Yan process ($0.13<x_2<0.45$)~\cite{dove2021,dove2023}.
For charmonium production, the gluon-gluon fusion
subprocess alone would give a cross section ratio
as $\left(1+ g_n(x_2)/g_p(x_2)\right)/2$,
where $g_{p,n}$ refers to the gluon distribution in the proton
or neutron. As the gluon is an iso-scalar particle, one expects
an identical charmonium production cross
section per nucleon for $p+d$ and $p+p$. This prediction clearly would
be modified once the contribution from the $q \bar{q}$ annihilation
subprocess to the charmonium production is included.
It should also be noted as the strong interaction is insensitive to the electric charge
of the quarks, the relative weighting between $u\bar{u}$ and $d\bar{d}$ is different
between charmonium production and the Drell-Yan process.
As a result, the charmonium production is less sensitive to the light sea-quark asymmetry than the Drell-Yan process.
The red solid curve in \cref{fig:pdpp}, obtained from the NRQCD calculation
using the NNPDF4.0 proton PDFs, is in good agreement with the $J/\psi$
cross section ratio data.
The clear deviation from unity for the calculated
ratio indicates a sizable contribution from the $q \bar{q}$ annihilation
at the large $x_F$ region,
even though the gluon-gluon fusion remains important for $J/\psi$ production
at lower $x_F$. For comparison, the black dot-dashed curve
in \cref{fig:pdpp}, corresponding to the NLO calculation for the Drell-Yan cross
section ratio, gives significantly larger values for the ratio in qualitative
agreement with the data.
Our results are also compared with CEM calculations in \cref{fig:pdpp_CEM} of the Supplemental Material.

In summary, the SeaQuest experiment has measured
the cross sections for $J/\psi$ and $\psi\left(2S\right)$ in $p+p$ and
$p+d$ interactions at \SI{120}{\GeV}. The $x_F$ dependence of the
$J/\psi$ and $\psi\left(2S\right)$ production cross sections is well described by
the NRQCD calculation. The $\sigma_{\psi\left(2S\right)}/\sigma_{J/\psi}$ ratio
is also shown. The measured ratio increases as $x_F$ increases, indicating
the increasing importance of $q\bar{q}$ annihilation in $\psi\left(2S\right)$ production.
The $p_T$ dependence is also reported. The extracted $\expval{p_T^2}$ from
this measurement follows an increasing pattern versus $\sqrt{s}$ established
by data over a wide range of energies.

We also present a direct comparison of $\sigma_{pd}/2\sigma_{pp}$
between $J/\psi$ production and the Drell-Yan process. While the
Drell-Yan process proceeds
via $q \bar{q}$ annihilation, $J/\psi$ production has contributions from
both the $q \bar{q}$ annihilation and the $g g$ fusion processes. The measured
$\sigma_{pd}/2\sigma_{pp}$ ratios are greater than unity for both the
Drell-Yan and $J/\psi$ production, showing that both processes are
sensitive to the $\bar{d}, \bar{u}$ flavor asymmetry of the proton sea.
The smaller values of $\sigma_{pd}/2\sigma_{pp}$ for $J/\psi$
production reflect the dilution due to the additional contribution
of $g g$ fusion for charmonium production. It would be interesting to
include the $\sigma_{pd}/2\sigma_{pp}$ $J/\psi$ data in a future
extraction of the $\bar{d}/ \bar{u}$ asymmetry of the proton.

\section*{Acknowledgments}
We thank G. T. Garvey for contributions to the early stages of this experiment
and M. Diefenthaler, L. Guo, B. Kerns, N. Kitts, N. C. R. Makins, I. Mooney,
A. J. Puckett, D. Su, B. G. Tice, S. Watson, Z. Xi for their contributions to the execution of the experiment. We also thank the Fermilab Accelerator Division and Particle Physics Division for their support of this experiment. This work was performed by the SeaQuest Collaboration, whose work was supported in part by the
US Department of Energy under Grant No.~DE-AC02-06CH11357, No.~DE‐FG02‐03ER41243, No.~DE-FG02-07ER41528, and No.~DE-FG02-94ER40847; the US National Science Foundation under Grants No.~PHY 2013002, No.~PHY 1812340, No.~PHY 2110898, No.~PHY 2110229, No.~PHY 2111046, No.~PHY 2209348, and No.~PHY 2309922; National Science and Technology Council of Taiwan (R.O.C.); the JSPS (Japan) KAKENHI through Grants No.~21244028, No.~25247037, No.~25800133, No.~20K04000 and No.~22H01244. Fermilab is managed by Fermi Research Alliance, LLC (FRA), acting under Contract No.~DE-AC02-07CH11359. 

\bibliographystyle{elsarticle-num}

\clearpage

\setcounter{figure}{0}
\setcounter{section}{0}
\renewcommand{\theequation}{S\arabic{equation}}
\renewcommand{\thefigure}{S\arabic{figure}}
\renewcommand{\thetable}{S\arabic{table}}
\renewcommand{\thesection}{S-\Roman{section}}
\section*{Supplemental Material}
\begin{figure}[ht!]
	\includegraphics[width=\linewidth]{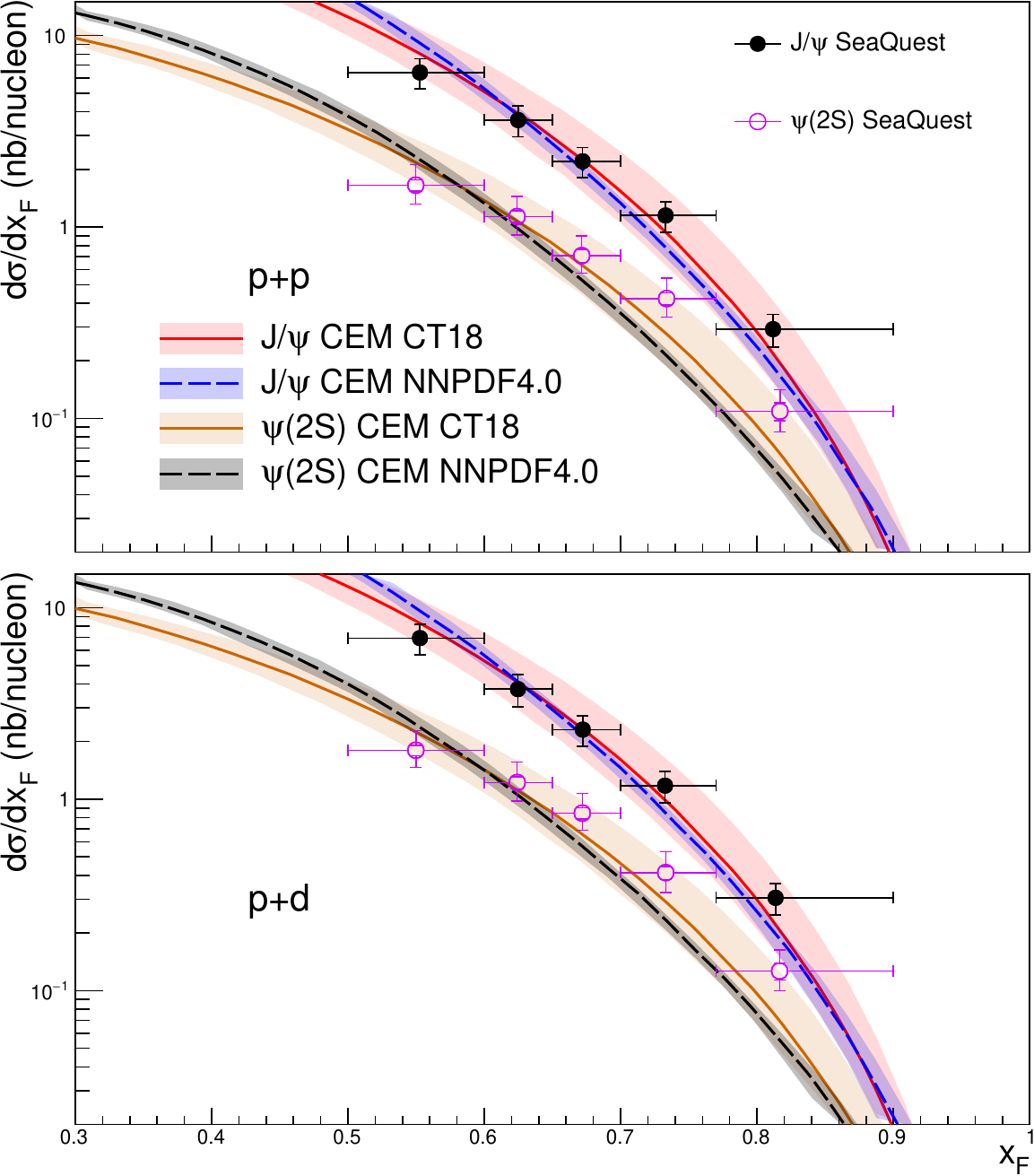}
	\caption{
Same as \cref{fig:cs_xF}, except that the curves correspond to CEM calculations~\cite{nason1988,nason1989,mangano1993,schuler1996,nelson2013}
		using the nucleon PDFs from CT18~\cite{hou2021} and NNPDF4.0~\cite{ball2022a}.
		The normalizations are obtained by fitting to the data.
		The error bands indicate 68\% confidence level from the PDFs.}
	\label{fig:cs_xF_CEM}
\end{figure}

\Cref{fig:cs_xF_CEM} shows the comparison of the extracted charmonium cross section with calculations using CEM~\cite{nason1988,nason1989,mangano1993,schuler1996,nelson2013}
and nucleon PDFs from CT18~\cite{hou2021} and NNPDF4.0~\cite{ball2022a},
The normalization of the calculations is obtained from fitting to our measurements. 
The measured $J/\psi$ cross sections are in good agreement with the CEM calculations,
but the CEM calculations underestimate the $\psi\left(2S\right)$ at large $x_F$.
It should be noted that in the CEM framework, the hadronization probability is independent of the underlying sub-processes,
and only depends on the final charmonium states.
Hence, in CEM, the shape of the $x_F$ distributions are nearly identical for $J/\psi$ and $\psi(2S)$.
Consequently, CEM would predict the $\sigma_{\psi(2S)}/\sigma_{J/\psi}$ ratio to be largely independent of $x_F$.
\vfill\break

\begin{figure}[ht!]
	\includegraphics[width=\linewidth]{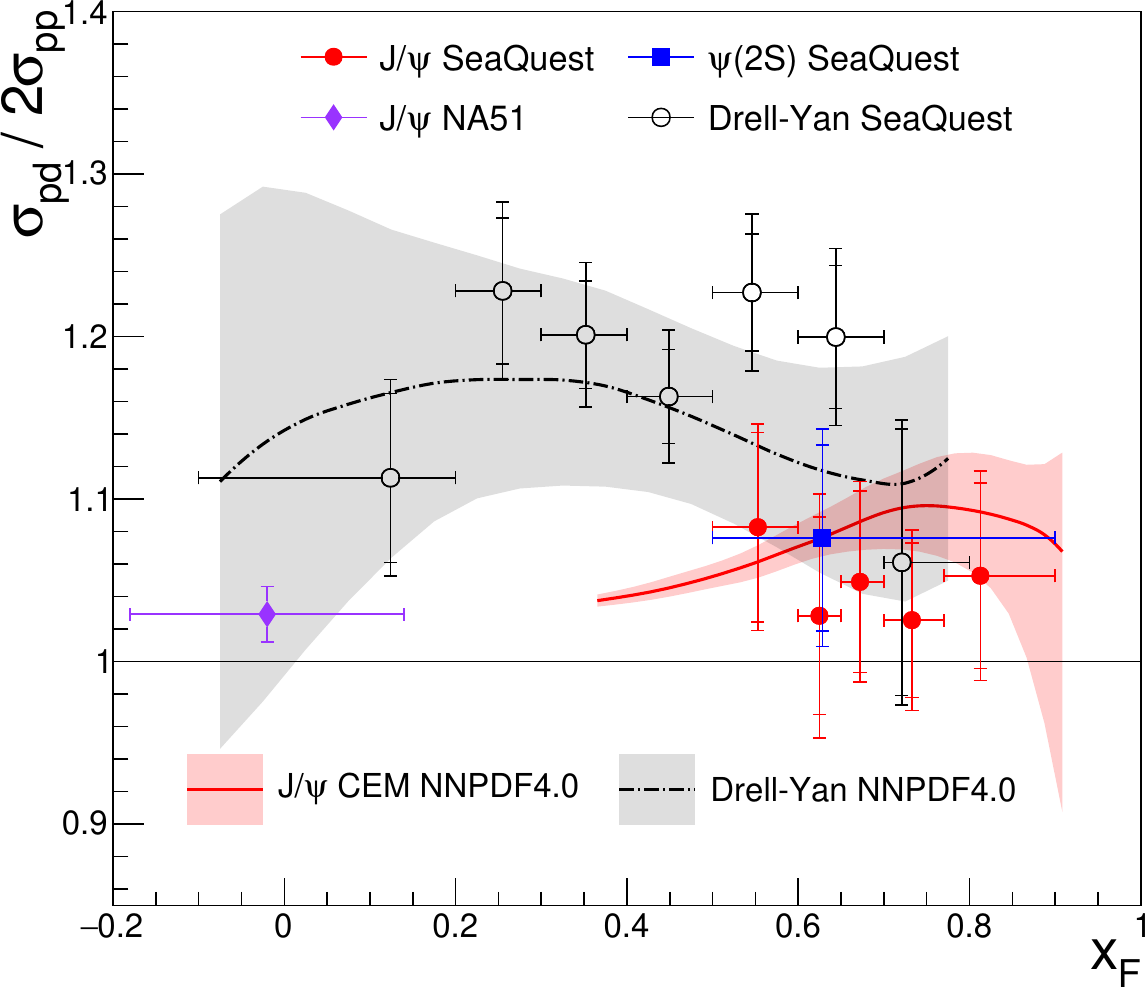}
	\caption{
Same as \cref{fig:pdpp}, except that the solid curve is the calculation for the charmonium cross section
		ratios at \SI{120}{\GeV} in the CEM using the proton PDFs
		from NNPDF4.0~\cite{ball2022a}.
The error bands indicate 68\% confidence level from the PDFs.}
	\label{fig:pdpp_CEM}
\end{figure}

The measured $\sigma_{pd}/2\sigma_{pp}$ cross section ratios for charmonium production are compared to CEM in \Cref{fig:pdpp_CEM}.
As both $J/\psi$ and $\psi(2S)$ share the same $x_F$ dependence,
the expected cross section ratios from CEM are nearly identical between the two charmonium states.
The measured cross section ratios are in agreement with the CEM calculations within experimental uncertainties.
\end{document}